# Magnetotransport as a probe of phase transformations in metallic antiferromagnets: the UIrSi$_3$ case

F. Honda[1], J. Valenta[2], J. Prokleška[2], J. Pospíšil[2], P. Proschek[2], J. Prchal[2], and V. Sechovský[2]

[1]*Tohoku University, Institute for Materials Research, Narita-cho 2145-2, Oarai, Ibaraki, Japan*
[2]*Charles University, Faculty of Mathematics and Physics, Department of Condensed Matter Physics, Ke Karlovu 5, Prague 2, Czech Republic*

**Abstract**

The electrical resistance, Hall resistance and thermoelectric power of the Ising-like antiferromagnet UIrSi$_3$ were measured as functions of temperature and magnetic field. We have observed that the unequivocally different characters of first-order and second-order magnetic phase transitions lead to distinctly different magnetotransport properties in the neighborhood of corresponding critical temperatures and magnetic fields, respectively. The magnetic contributions to the electrical and Hall resistivity in the antiferromagnetic state, and the polarized and normal regimes of paramagnetic state are driven by different underlying mechanisms. Results of detailed measurements of magnetotransport in the vicinity of the tricritical point reveal that the Hall-resistivity steps at phase transitions change polarity just at this point. The jumps in field dependences of specific heat, electrical resistivity, Hall resistivity and Seebeck coefficient at the first-order metamagnetic transitions indicate a Fermi surface reconstruction, which is characteristic of a magnetic-field induced Lifshitz transition. The presented results emphasize the usefulness of measurements of electrical- and thermal-transport properties as sensitive probes of magnetic phase transformations in antiferromagnets sometimes hardly detectable by other methods.

**Introduction**

Since the electrical transport can be influenced by interactions of conduction electrons with magnetic fields and with unpaired electrons carrying magnetic moments, the electrical resistivity and Hall resistivity may serve as important probes of details of magnetism in metallic materials.

The electrical resistivity $\rho$ in magnetic metals is considered within a simple approach, supposing validity of Mathiessens' rule, as a sum:

$$\rho = \rho_0 + \rho_{e-p} + \rho_{mag} \qquad (1).$$

The temperature independent residual-resistivity term $\rho_0$ which originates in the scattering of conduction electrons from lattice defects, and the electron-phonon term $\rho_{e-p}$ reflecting the scattering of conduction electrons from phonons are present in all metallic materials. The latter term represents the scattering of conduction electrons from magnetic moments due to exchange interaction with unpaired electrons carrying the moment.

The states of 5f-electrons carrying magnetic moments in uranium intermetallics can form a narrow band at the Fermi level. The strong interaction with conduction electron states causes significantly enhanced scattering of conduction electrons from U magnetic moments. The $\rho_{mag}$ values of U intermetallics in a paramagnetic (PM) state are usually high and roughly temperature independent. At temperatures below the magnetic-ordering temperature, $\rho_{mag}$ decreases with temperature in a way characteristic for magnetic excitations, especially magnons[1,2].

In ferromagnets, $\rho_{mag}$ vanishes in the low temperature limit. On the other hand, rather large $\rho_{mag}$ values are usually observed for antiferromagnetic (AFM) U materials even at the lowest temperatures. Resistance measurements on anisotropic materials reveal the anomalously large $\rho_{mag}$ low-temperature values for the current applied along the directions with AFM coupling of magnetic moments[3,4].

The uranium-based antiferromagnets with uniaxial anisotropy[3,5-8] exhibit magnetic behavior like the Ising antiferromagnets[9,10]. These strongly anisotropic antiferromagnets are generally characterized by simple reversals of the local magnetic moment directions. When cooled in zero field, they undergo a second-order magnetic phase transition (SOMPT) from PM to an AFM state at $T_N$. Below $T_N$, they are ordered antiferromagnetically with a sublattice structure, in which the large anisotropy constrains the magnetic moments to point either parallel or antiparallel to the easy axis. On the application of a magnetic field along the easy axis a metamagnetic transition (MT) from the AFM to the PM state takes place at a critical field $H_c$. At sufficiently low temperatures, MT is a first-order magnetic phase transition (FOMPT) characterized by a sudden reversal of antiparallel sublattices to the direction to the applied field. The high-field ($H > H_c$) state is characterized by ferromagnetic-like aligned magnetic moments but it is a paramagnetic (not ferromagnetic) state[11]. Due to its character it is called a field polarized paramagnet (PPM) regime[12,13].

The first-order metamagnetic transition between the AFM state and the PPM regime in uranium-based antiferromagnets with uniaxial anisotropy is accompanied by a dramatic drop of electrical resistivity[3,5-8,14,15] and $\rho_{mag}$ practically vanishes in the low-temperature limit similar to $\rho_{\text{mag}}$ in ferromagnets. The negative magnetoresistance jumps observed at $H_c$ on uranium-based antiferromagnets[3,5-8,14,15] quantitatively compare to the giant magnetoresistance (GMR) reported in magnetic multilayers[16].

The large low-temperature resistivity and negative magnetoresistance values, respectively, in antiferromagnets have generally two underlying mechanisms. When the AFM periodicity does not coincide with the crystallographic (chemical) unit cell, a reconstruction of the Fermi surface may occur at the transition temperature, assuming that a new Brillouin zone boundary cuts the Fermi surface. As a result, an electron energy gap may be created along the new periodicity direction. This leads to a reduction of effective number of charge-carriers and a consequent increase of resistivity. This approach has been used for explanation of the $\rho$ increase in AFM lanthanide compounds below $T_N$[17]. On the other hand, the AFM periodicity is removed by the MT from AFM to PPM, so the AFM gaps in Fermi surface are consequently closed and the electrical conductivity is correspondingly recovered.

The major portion of the giant negative magnetoresistance observed at the MT in the Ising-like uranium antiferromagnets cannot be explained by the mechanism based on a Fermi surface gapping. The spin-dependent scattering mechanism involving mainly the scattering due to ↑↓ coupled U magnetic moments in the AFM structure[4] may be considered as dominant. This concept is analogous to the approach to GMR in magnetic multilayers[16,18,19]. Within a certain interval of temperatures below $T_N$ the MT is a continuous transition (SOMPT). The applied magnetic field ($H < H_c$) along the easy axis induces fluctuations from an AFM state. These fluctuations are multiplying with increasing magnetic field up to $H_c$. The conduction electrons scatter from the fluctuations that leads to progressively increasing $\rho_{\text{mag}}$[20]. At $H_c$, a peak in the $\rho_{\text{mag}}(T)$ dependence has been found by calculations[21] and experiment, e.g. on $V_5S_8$[22].

The line of critical points of the SOMPTs at high temperatures and the line of FOMPTs critical points at low temperatures meet at a point which is known as a tricritical point (TCP). No systematic magnetotransport data on anisotropic U antiferromagnets involving SOMPTs and their evolution in the vicinity of a TCP have been reported so far.

The ordinary Hall effect arising from the Lorentz force acting on the charge carriers turned out to be a useful tool for determination of charge-carrier density in nonmagnetic materials and played an important role in the early-years of semiconductor physics research as well as related solid-state electronics. The normal Hall resistivity provides, for single-band metals, a measure of the volume in momentum space enclosed by the Fermi surface. In materials possessing magnetization an additional contribution comes into play as a consequence of the anomalous Hall effect (AHE). The total Hall resistivity can be described empirically as a sum of two terms; the normal and the anomalous Hall resistivity[23,24,25-27]:

$$\rho_H(H) = R_H \cdot \mu_0 H = R_o \cdot \mu_0 H + R_s \cdot M \qquad (2),$$

where $R_o$ and $R_s$ are the normal and the anomalous Hall coefficient, respectively, $H$ is the applied magnetic field and $M$ is the volume magnetization both perpendicular to the plane in which the Hall resistivity is measured. Thus the single Hall coefficient is written as:

$$R_H = R_o + R_s \cdot \frac{M}{\mu_0 H} \qquad (3)$$

An AHE is caused by three underlying mechanisms – the first one is intrinsically caused by specific features of band structure (Berry phase), the other two involve the left-right asymmetric scattering due to the skew scattering and the side-jump scattering of conduction electrons. The anomalous Hall coefficient can be expressed as a sum of two terms:

$$R_s = a \cdot \rho + b \cdot \rho^2 \quad (4),$$

where $a \cdot \rho$ represents the skew-scattering and $b \cdot \rho^2$ the intrinsic and side-scattering mechanisms. The AHE has been in fact recognized on ferromagnetic iron already by Hall[28]. Pugh and Lippert[23,24] have shown that the empirical formula (2) applies to many materials over a broad range of external magnetic fields. In ferromagnets, the second term represents the contribution due to the spontaneous magnetization[23-26]. The studies of Hall-effect in antiferromagnets have a much shorter history than the research of the AHE in ferromagnets. Recently, much interest has been paid to the AHE in noncollinear transition-metal antiferromagnets in which sizable AHE can be found also in the state with zero net magnetization[29-32]. These materials offer promising opportunities in topological antiferromagnetic spintronics[33].

The Hall effect was investigated in several antiferromagnetic *f*-electron intermetallics within the periods of research interest in fluctuating-valence, Kondo-lattice and heavy-fermion lanthanide[34-39] and uranium compounds[40-44]. Several papers were dedicated to investigation of the Hall effect related to metamagnetic transitions in AFM materials[45-50].

This paper is devoted to detailed investigation of magnetotransport properties of UIrSi$_3$ in relation with its specific magnetism. It is one of the only two known uranium intermetallic compounds adopting the non-centrosymmetric tetragonal BaNiSn$_3$-type structure. Antiferromagnetism of UIrSi$_3$ at temperatures below 42 K has been first reported by Buffat et al.[51] from experiments on polycrystals. They also observed a metamagnetic-like transition in $\mu_0 H$ = 5.6 and 3.2 T at $T$ = 30 and 38 K, respectively.

Recently, UIrSi$_3$ single crystals have been grown and subjected to magnetization and specific-heat measurements[6]. The antiferromagnetism below the Néel temperature $T_N$ = 41.7 K has been confirmed. Magnetization and specific-heat data revealed a strong uniaxial anisotropy in the AFM state with the *c*-axis as the easy magnetization direction which places UIrSi$_3$ among the Ising systems. When a magnetic field is applied along the *c*-axis it undergoes a MT from the AFM to a PM state at a critical field $H_c$. No MT is observed when the field is applied along the *a*-axis, up to 14 T.

At temperatures below 28 K, the MT is a FOMPT ($\mu_0 H_c$ = 7.3 T at 2 K) to a PPM regime. The saturated magnetization in the PPM regime amounts to 0.66 $\mu_B$/f.u. This value is rather small in comparison to the expected values of the $U^{3+}$ and $U^{4+}$ free-ion ordered moments, 3.20 and 3.27 $\mu_B$, respectively, which suggests an itinerant character of the 5f-electron magnetism (if Ir and Si magnetic moments can be neglected).

A second-order metamagnetic transition is observed at higher temperatures (28 K > $T$ > $T_N$). The point in the *H-T* magnetic phase diagram where the transition switches between a FOMPT and a SOMPT is considered as the tricritical point (at $T_{tcp}$ = 28 K and $\mu_0 H_{tcp}$ = 5.8 T).

The main objective of the present work is to determine the manifestation of various magnetic phase transitions in the Ising itinerant 5f-electron antiferromagnet UIrSi$_3$ in magneto-transport properties. For this purpose numerous isofield $\rho_{[001]}(T)$, $\rho_{[100]}(T)$, $\rho_H(T)$ and isothermal $\rho_{[001]}(H)$, $\rho_{[100]}(H)$, $\rho_H(H)$ dependences were measured within wide intervals of temperatures (2 – 300 K) and fields (0 – 14 T) parallel to the easy-magnetization direction, i.e. *c*-axis. To assure the best quality of samples a new UIrSi$_3$ single crystal has been grown employing our experience from previous work[6].

All three resistivities were found to be sensitive to magnetic-phase transitions in UIrSi$_3$. The $\rho(T)$ and $\rho_H(T)$ dependences measured in various magnetic fields exhibit considerable anomalies at corresponding critical temperatures $T_N(H)$. The $\rho(H)$ and $\rho_H(H)$ isotherms show anomalies at corresponding critical fields of MT, $H_c(T)$. The $T_N(H)$ and $H_c(T)$ values fit very well with the magnetic phase diagram[6] derived using magnetization and specific-heat measurements. The character of the anomalies corresponding to FOMPTs and SOMPTs has been found strikingly different. The observed change of polarity of the $\Delta\rho_H(T)$ and $\Delta\rho_H(H)$ steps at the temperature and magnetic field where the FOMPT changes to a SOMPT may offer a useful criterion for determination of the tricritical point (TCP) in Ising antiferromagnets.

We have also measured the thermoelectric effect at several temperatures as a function of magnetic field. The drop of the value of the Seebeck coefficient observed at the $H_c$ of the FOMPT in conjunction with the corresponding jumps in $\rho(H)$, $\rho_H(H)$ and $C_p(H)$ dependences provide strong indications that the FOMPT in UIrSi$_3$ is probably a Lifshitz transition, which is characterized by a Fermi surface reconstruction.

**Experimental**

A UIrSi$_3$ single crystal has been prepared by the floating zone melting method in a commercial four-mirror optical furnace with halogen lamps, each 1kW (modelFZ-T-4000-VPM-PC, Crystal Systems Corp., Japan). In the first step, a polycrystalline material of UIrSi$_3$ was synthesized by arc-melting from stoichiometric amounts of the pure elements U (3N, further treated by Solid State Electrotransport[52,53]), Ir (4N), and Si (6N) in Ar (6N) protective atmosphere. No sign of evaporation was detected during the melting. Then, a precursor in the form of a 50 mm long rod was prepared by arc melting in a special water-cooled copper mold at identical protective conditions. The quartz chamber of the optical furnace was evacuated by a turbomolecular pump to $10^{-6}$ mbar before the crystal growth process. In order to desorb gases from the surface of the precursor, the power of the furnace was increased gradually up to 30% of maximum power (far below the melting at ~54% power) and the precursor was passed through the hot zone several times while continuously evacuating. After the degas process and evacuation, the quartz chamber was quickly filled with high purity Ar (6N). The whole growth process was performed with Ar flow of 0.25 l/min and a pressure of ~2 bar. A narrow neck was created in the beginning of the growth process by variation of the speed of the upper and bottom pulling shafts. The pulling rate was very slow, only 0.5 mm/h, and without rotation. A large single crystal of the cylindrical shape with length ~50 mm and diameter 4 mm was obtained. The high quality and orientation of the single crystal was verified by the Laue method (Fig. 1). The stoichiometric composition was verified by a scanning electron microscopy (SEM) using a Tescan Mira I LMH system equipped with an energy-dispersive X-ray detector (EDX) Bruker AXS. The analysis revealed a single phase single crystal of 1:1:3 composition. Detailed surface analysis did not detect any foreign phases.

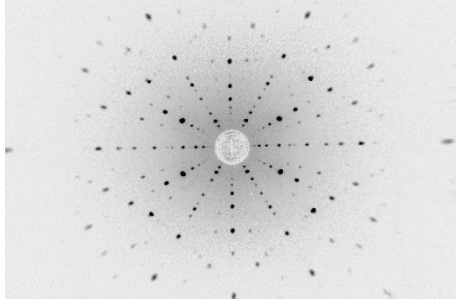

Fig. 1. Laue pattern of the UIrSi$_3$ single crystal oriented along [110].

First characterization of crystal has been done by magnetization and specific-heat measurements analogous to these reported in ref. 6. The obtained results were in fair agreement with data presented before[6].

All data presented in this paper have been measured in magnetic fields applied exclusively along the *c*-axis of the tetragonal structure of UIrSi$_3$. The electrical resistivity, Hall resistivity, thermoelectric power, magnetization and specific heat were measured with a physical property measurement system (PPMS, Quantum Design Inc.) in fields up to 14 T. For determination of $T_N$ from the temperature dependence of the specific heat, the point of the balance of entropy released at the phase transition method was used. The specific heat was measured on a basal-plane plate sample of 11 mg mass. Resistivity measurements were performed on two bar-shape samples (1.8×0.75×0.73 mm$^3$ and 1.1×0.78×0.55 mm$^3$ for current applied along the *a*- and *c*-axis, respectively). The Hall resistivity was measured with a basal-plane plate sample (diameter of 1.2 mm) with current applied along the *a*-axis and the Hall voltage measured in the perpendicular direction in the basal plane. The sample for thermoelectric power measurements was a 1×1×4 mm$^3$ *c*-axis bar.

The field dependences of electrical resistivity $\rho(H)$, Hall resistivity $\rho_H(H)$, thermoelectric power $S(H)$ and magnetization $M(H)$ were measured in fields between 4 and 8 T at a sweep rate of 1 mT/s, 2.5 mT/s, 2.5 mT/s and 2 mT/s, respectively. The system has been found only slightly relaxing. A typical time dependence of electrical resistance at a most "sensitive" point of the hysteresis loop (see Fig. S1 in Supplementary information) is seen in Fig. S2. This demonstrates that the hysteresis of the FOMPT observed in $\rho(H)$, $\rho_H(H)$, $S(H)$ and $M(H)$ is intrinsic and not any artefact of fast sweeping the applied magnetic field.

**Results and Discussion**

The observed anisotropy of the temperature dependence of the electrical resistivity, $\rho(T)$ (see Fig. 2) indicates an anisotropic Fermi surface of UIrSi$_3$. The resistivities $\rho_{[100]}(T)$ and $\rho_{[001]}(T)$ for current *i* // [100] and [001], respectively, increase with increasing temperature above $T_N$ and gradually saturate (the curvature and tendency to saturation is more pronounced in the $\rho_{[001]}(T)$ dependence). This resembles the behavior of transition metals and their compounds characterized by a narrow *d*-electron band crossing the Fermi level ($E_F$), which was explained by an *s-d* scattering mechanism as proposed by Mott[54] and Jones[55]. We tentatively suppose that the resistivity of U intermetallics characterized by a narrow 5f-electron band crossing the $E_F$ could be considered within an analogous s-*f* scattering model.

The negative curvature of both, $\rho_{[100]}(T)$ and $\rho_{[001]}(T)$ observed at high temperatures suddenly changes to a convex dependence at the same characteristic temperature, which coincides with the $T_N$ value determined from specific-heat data (see inset of Fig. 2). The RRR values are 34 and 14, respectively.

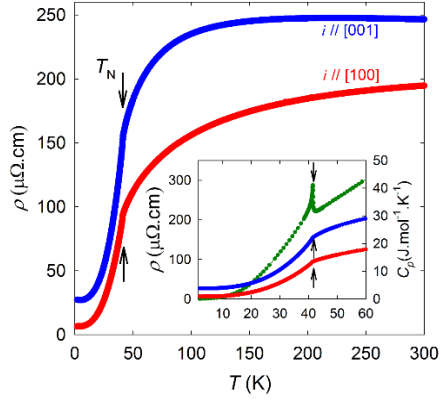

Fig. 2. Temperature dependence of the electrical resistivity of UIrSi$_3$ for electrical current parallel to the [100] and [001] direction, respectively. Inset: a low-temperature detail including also the corresponding specific-heat $C_p$ vs. $T$ (green points) plot. The arrows marks $T_N$.

When we apply the magnetic field along the [001] direction the $T_N$-related anomaly in the $\rho_{[100]}(T)$ and $\rho_{[001]}(T)$ dependences are shifted to lower temperatures with increasing field such that they follow the corresponding specific-heat anomaly (see Fig. 3). The resistivity anomaly at $T_N$ simultaneously develops with increasing the field from a just-negative $\partial\rho/\partial T$ change in zero field to a clear positive $\Delta\rho$ step in 5 T for the AFM to PM transition. $T_N$ is associated with the maximum of $\partial\rho/\partial T$. In 6 T we suddenly observe a negative $\Delta\rho$ step at $T_N$ for the AFM to PPM transition, which is evidenced also by the zero-field-cooled (ZFC) curves measured in 7 T. The 8-T $\rho_{[100]}(T)$ and $\rho_{[001]}(T)$ curves are smooth showing no sharp anomaly within the entire temperature range. A detailed view of the evolution of $\rho_{[001]}(T)$ curves in fields from 5 to 8 T is displayed in Fig. 4. The corresponding magnetization $M(T)$ dependences shown in the same figure exhibit a positive $\Delta M$ step at $T_N$ in the fields of 5 and 6 T, respectively, which is followed by a decay of the magnetization with further increasing temperature. The $T_N$-related anomalies in the $\rho_{[001]}(T)$ and $M(T)$ curves measured in 6 T exhibit a temperature hysteresis, which is characteristic for a first-order phase transition. In contrast, the $T_N$-related anomaly in fields up to 5 T show no hysteresis.

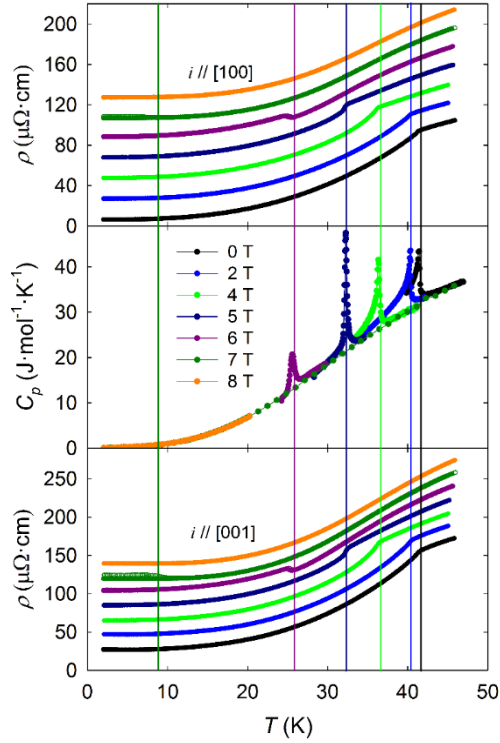

Fig. 3. Temperature dependences of the electrical resistivity for the current parallel to the [100] and [001] direction (top and bottom panel, respectively) and specific heat (middle panel) of UIrSi$_3$ below 45 K in the magnetic field applied in the [001] direction. The $\rho$ vs. $T$ curves measured in different fields are mutually shifted by 20 µΩ·cm along the vertical axis for clarity. The actual vertical scale corresponds to the 0-T curve. The colored vertical lines represent the $T_N$ values corresponding to the actual applied magnetic fields. The 7-T line corresponds to the bifurcation point of the FC and ZFC resistivity curves.

The Hall resistivity, $\rho_H$, in field parallel to [001] is also sensitive to the PM ↔ AFM transition at $T_N$ as can be seen in Fig. 5. The $T_N$-related anomaly in the $\rho_H(T)$ corresponding to a gradually increasing magnetic field undergoes a development analogous to the normal-resistivity case. It is gradually shifted to lower temperatures in coincidence with the $T_N$-related specific-heat and magnetization anomalies. The Hall resistivity anomaly simultaneously develops with increasing the field from a positive $\partial \rho_H/\partial T$ change in 1 T to a clear negative $\Delta \rho_H$ step in 5 T. $T_N$ coincides with the minimum of $\partial \rho_H/\partial T$ and roughly with the maximum of $\partial M/\partial T$. Also, the Hall resistivity exhibits a contrast between the $T_N$-related anomalies in fields up to 5 T and those measured in higher fields. In 6 T we suddenly observe a positive $\Delta \rho_H$ step at $T_N$ with temperature hysteresis. The observed qualitative changes of the $T_N$-related anomalies in the corresponding $M(T)$, $\rho_{[100]}(T)$, $\rho_{[001]}(T)$ and $\rho_H(T)$ dependences in fields between 5 and 6 T may be considered in connection with the conclusion in Ref. 6, that the change from a SOMPT to a FOMPT happens at a TCP which has been estimated at $\mu_0 H_{tc}$ ~ 5.8 T, $T_{tcp}$ ~ 28 K.

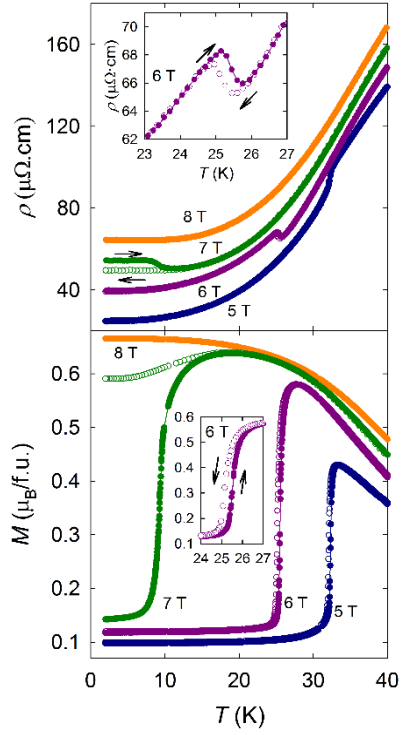

Fig. 4. Temperature dependences of the electrical resistivity $\rho_{[001]}$ (upper panel) and magnetization *M* (lower panel) of UIrSi$_3$ measured in the magnetic field of 5, 6, 7 and 8 T, respectively, applied in the [001] direction. For 7 T the ZFC (line with open symbols) and FC (line with full symbols) *M(T)* and $\rho$*(T)* curves, respectively, bifurcate below $T_N$. The $\rho$*(T)* curves measured in different fields are mutually shifted by 15 µΩcm along the vertical axis for clarity. The displayed vertical scale corresponds to the 5-T curve. Inset of lower panel: detail of the hysteresis of the transition in 6 T. The arrows represent the direction of field sweep.

The ZFC $\rho_H(T)$ curve measured in 7 T also shows a step, which is however considerably larger. Similar to normal resistivity and magnetization behavior, the 8-T $\rho_H(T)$ curve is smooth showing no sharp anomaly within the entire temperature range. The 8-T field is sufficiently higher than $\mu_0 H_c$ (7.3 T) at 2 K[6] to entirely destroy the AFM ordering in the ZFC sample and recover the PM state (PPM at sufficiently low temperatures). Application of an 8-T field, when cooling UIrSi$_3$ from high temperatures, prevents any transition to the AFM ordering, i.e. the sample remains PM (PPM at low *T*). That is why the corresponding 8-T field-cooled (FC) and ZFC *M(T)*, $\rho_{[100]}(T)$, $\rho_{[001]}(T)$, $\rho_H(T)$ curves, respectively, are identical and exhibit no $T_N$-related anomaly (see Figs. 4 and 5).

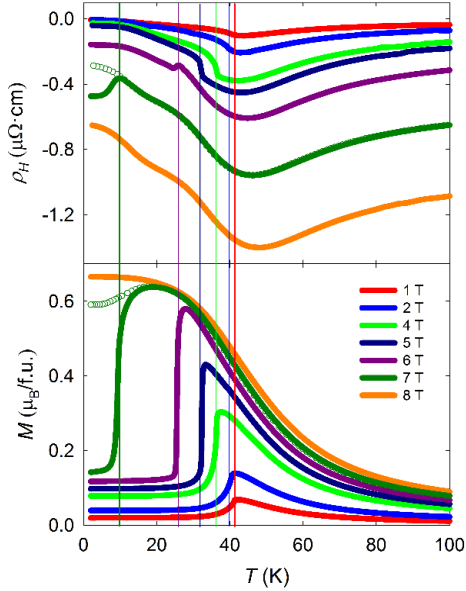

Fig. 5. Temperature dependence of Hall resistivity $\rho_H$ vs. $T$ (upper panel) and magnetization $M$ vs. $T$ (lower panel) of UIrSi$_3$ in several magnetic fields applied in the [001] direction. The $\rho_H$ vs. $T$ and $M$ vs. $T$ plots in corresponding magnetic fields are in the same colors. The colored vertical lines represent the $T_N$ values determined by specific-heat measurements. The 7-T vertical line corresponds to the bifurcation point of the ZFC (line with full symbols) and FC (line with open symbols) $\rho_H$ vs. $T$ and $M$ vs. $T$ curves, respectively. The 6-T, 7-T and 8-T plots in the upper panel are vertically shifted by - 0.1, - 0.4, - 0.8 μΩcm, respectively.

The corresponding ZFC and FC $M(T)$, $\rho_{[100]}(T)$, $\rho_{[001]}(T)$, $\rho_H(T)$ curves measured in 7 T bifurcate in the vicinity of $T_N$ (see Figs. 4 and 5). This is reflecting the large field hysteresis of the MT reported in Ref. 6 which extends around 7 T at low temperatures. When cooling UIrSi$_3$ in 7 T the $M(T)$ values reach a maximum at ~ 20 K then decrease by about 10 % on further cooling. This indicates that the low-temperature FC state is a somewhat disturbed PPM which, however, exhibits considerably lower resistivity than the low-temperature ZFC state (probably the AFM ground state). The higher resistivity in the AFM state can be also due to the Fermi surface truncated by energy gaps caused by a different periodicity of the crystallographic and AFM lattices.

The entire $\rho_H(T)$ dependences measured between 2 and 100 K in fields up to 8 T (see Fig. 5) show a broad valley. The temperature of its minimum roughly coincides with the temperature of the $\partial M/\partial T$ minimum. The $\rho_H$ values are negative, as expected for the ordinary Hall effect in case of electron conductivity in metals. The positive values in the 8-T and FC 7-T $\rho_H(T)$ dependences at low temperatures reflect large positive contributions due to the AHE in UIrSi$_3$ in the PPM state.

The $\rho_{[100]}(H)$ and $\rho_{[001]}(H)$ data collected at selected temperatures shown in Fig. 6 demonstrate the evolution of MT related resistivity anomalies. In the lower panels, results obtained at $T < T_{tcp}$ at which a FOMPT takes place are displayed. The $\rho_{[001]}(H)$ and $\rho_{[001]}(H)$ curves in the vicinity of $H_c$ qualitatively resemble the corresponding magnetization curves in Ref. 6 taken with negative sign, i.e. the resistivity sharply drops at $H_c$ when sweeping the magnetic field up, and exhibits the asymmetric hysteresis of a MT when sweeping the field down.

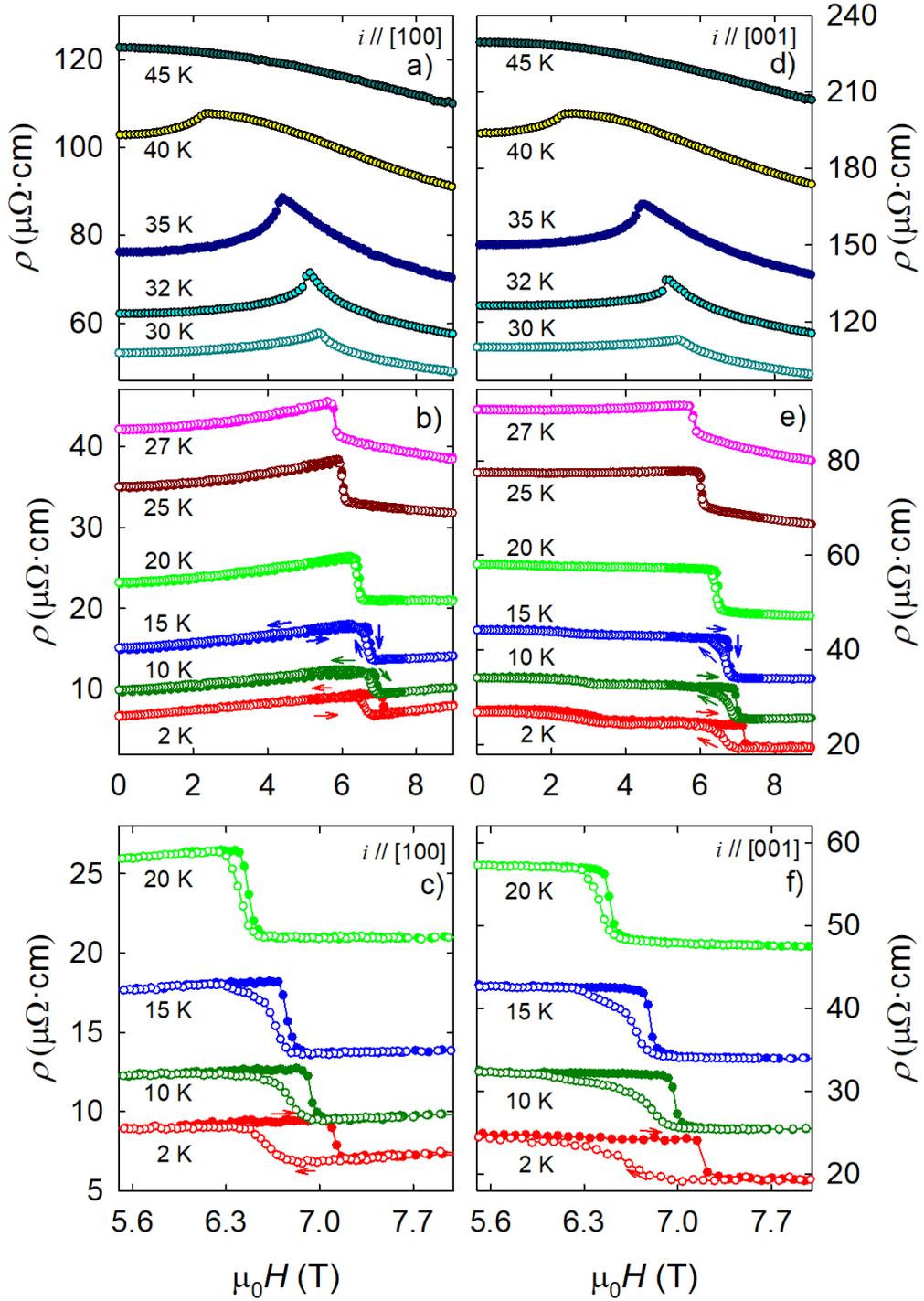

Fig. 6. The electrical resistivity of UIrSi$_3$ at selected temperatures for current parallel to the [100] (left panels a, b, c) and [001] (right panels d, e, f) direction as a function of the magnetic field applied in the [001] direction. The $\rho(H)$ curves in left (right) panels measured at different temperatures are mutually shifted by 2 μΩ·cm (6 μΩ·cm) along the vertical axis for clarity. The field scale of panels c and f is expanded to make the evolution of hysteresis at temperatures up to 20 K more visible. The arrows show the direction of field sweep.

The $\rho_{[100]}(H)$, $\rho_{[001]}(H)$ curves in the upper panels of Fig. 6 were measured at temperatures between $T_{tcp}$ and $T_N$. At these temperatures UIrSi$_3$ undergoes a field-induced SOMPT (AFM↔PM). A dramatic difference in the electrical resistivity response in comparison to the lower-temperature's FOMPT is clearly seen. Here the resistivity considerably increases with increasing field up to the maximum value $\rho(H_c)$. In fields beyond $H_c$ the resistivity values decay fast with increasing $H$ yielding a negative magnetoresistance well above $H_c$. Contrary to FOMPTs, these transitions have no hysteresis.

The Hall-resistivity isotherms $\rho_H(H)$ measured at temperatures below 28 K show a sudden positive $\Delta\rho_H(H)$ step at $H_c$ and an asymmetric hysteresis, being at lowest temperatures very similar to the magnetization behavior around the FOMPT at $H_c$[6]. In contrast, the $\rho_H(H)$ curves measured at temperatures higher than 28 K exhibit a slightly rounded negative step at $H_c$ and no field hysteresis. The step gradually smears out with increasing temperature to disappear at temperatures around 40 K. Note that $\Delta\rho_H(H)$ decreases (but does not scale) with the decreasing corresponding $\Delta M(H)$ step.

The observed opposite polarity of the Hall effect step accompanying the FOMPT and the SOMPT, respectively, points to a possible criterion for determination of the TCP, which separates the FOMPT and SOMPT sections of the magnetic phase diagram.

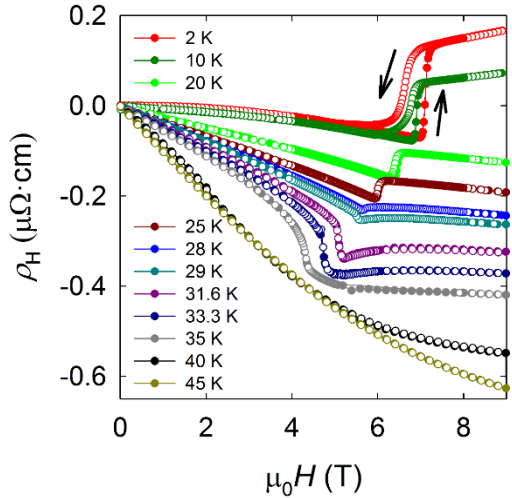

Fig. 7. The Hall resistivity of UIrSi$_3$ at selected temperatures as a function of the magnetic field applied in the [001] direction. Where needed, the arrows show the direction of field sweeps. The hysteresis of the MT at 20 K is 0.25 T, negligible at 25 K and zero at temperatures $\geq$ 28 K.

The main objective of the present study is determination and understanding of the impact of the SOMPT and the FOMPT in UIrSi$_3$ on magnetotransport properties. This would contribute to usability of magnetotransport as a probe of the type of magnetic phase transitions of antiferromagnets. Closer inspection of $M(T)$, $\rho_{[001]}(T)$ and $\rho_H(T)$ data (see Figs. 4 and 5) measured in the magnetic field parallel to $c$-axis reveals an evolution of magnetization of UIrSi$_3$ with cooling and heating, and the corresponding impact on magnetotransport. When cooling the crystal in a field of 8 T from high temperatures down to 2 K, UIrSi$_3$ is all the time in paramagnetic state. At 2 K, the highest $M$ and $\rho_H$ values, respectively, are recorded whereas $\rho$ reaches the lowest value. The magnetization is saturated, and so all magnetic moments are aligned (polarized) in the direction of the applied magnetic field, i.e. UIrSi$_3$ is in PPM regime. The same extreme values of $M$, $\rho_{[001]}$ and $\rho_H$, respectively, were measured after cooling the crystal in zero field down to 2 K and subsequently the field was applied and increased up to 8 T (see Fig. 6 in ref.6 and Figs. 6, 7 in this paper). After cooling in zero field to 2 K, UIrSi$_3$ appears in the AFM ground state. When a magnetic field is applied and increasing to 8 T, UIrSi$_3$ undergoes, at $H_c$, a FOMPT from the AFM to a PM phase with polarized magnetic moments, i.e. the PPM regime. The impact on magnetotransport is in both cases identical; $\rho_H$ reaches a maximum value and $\rho$ approaches a minimum. The positive step of the 2-K magnetization curve, due to the MT at $H_c$, is accompanied by a positive step of the $\rho_H(H)$ and a negative step of the $\rho(H)$ dependence.

In Figs. 6 and 7 (considering also Fig. 6 in ref.6), we can see that the positive step of the $\rho_H(H)$ and a negative one of the $\rho(H)$ dependence is observed at $T < 28$ K ($T_{tcp}$) for the transition from an AFM to a PPM, i.e. at which we observe the FOMPT at $H_c$. Strikingly different $\rho_H(H)$ and $\rho(H)$ behavior is observed at temperatures between $T_{tcp}$ and $T_N$ were UIrSi$_3$ undergoes a SOMPT for the transition from an AFM to a PM state.

We analyzed the Hall resistance data in detail within the scheme based on the empirical formulas (2, 3, 4) following from numerous investigations of the AHE in ferromagnets. In this course we fitted the isofield $\rho_H(T)$ and isothermal $\rho_H(H)$ data series to formulas (S1) and (S2), respectively (see Supplementary Information) in the context of the available magnetization and electrical resistance data. We have included representative results with descriptions in Supplementary Information.

The individual $\rho_H(T)$ and $\rho_H(H)$ dependences for different fields and temperatures can be reasonably formally fitted to the formulas, however, the variation of fitting parameters does not to have some physical background. Especially, no reasonable series of fits can be obtained for any chosen constant value of the ordinary Hall-effect coefficient $R_0$. A possible variation of $R_0$ in an itinerant 5f-electron antiferromagnet as UIrSi$_3$ might be due some reorganization the Fermi surface induced in the AFM state by magnetic fields considerably lower than $H_c$. Relevant band structure calculations may provide results collaborating this idea. In any case the scenario of the Hall effect in UIrSi$_3$ is most probably more complex than that usually investigated using the empirical approach of the AHE applied in case of ferromagnets.

At this stage of understanding we propose the following simple approach to explain the experimental findings:

The FOMPTs in UIrSi$_3$ are AFM ↔ PPM transitions, whereas the SOMPTs are AFM ↔ PM, where PM stands for a normal paramagnetic state, with normal thermal fluctuations of magnetic moments. The PPM regime at low temperatures, which is characterized by magnetic moments aligned along the field direction, resembles a ferromagnetic state. In the case of full polarization, it yields zero contribution to electrical resistivity with, on the other hand, a large contribution to anomalous Hall resistivity. The TCP separates the FOMPT and SOMPT regions in the magnetic phase diagram. The relation between the characteristic values of electrical resistance and anomalous Hall resistance of the three states (regimes) are:

$$\rho^{PM} > \rho^{AFM} > \rho^{PPM} \tag{5},$$

$$\rho_H^{PM} < \rho_H^{AFM} < \rho_H^{PPM} \qquad (6),$$

respectively.

In order to explore the details of the evolution of $\rho_H$ and $\rho_{[100]}$ anomalies in the neighborhood of the TCP in the *T-H* phase space, we performed thorough measurements of $\rho_H(T)$ and $\rho_{[100]}(T)$ isofield curves for fields 5.2, 5.3, …5.9, 6.0 T and ($\rho_H(H)$ and $\rho_{[100]}(H)$ isotherms at temperatures 25, 26, …, 30, 31 K. The results of these measurements are displayed in Figs. S15 and S16 in Supplementary Information. It is evident that $\Delta\rho_H(T)$ and $\Delta\rho_H(H)$ continually develop from positive to negative values with decreasing magnetic field and increasing temperature, respectively. Considering the estimated values of temperatures and fields for which $\Delta\rho_H(T)$ and $\Delta\rho_H(H)$ values pass through zero we conclude that the change of polarity of jumps of the AHE, as functions of temperature and magnetic field, take place at the TCP.

Closer inspection of isofield $\rho_{[100]}(T)$ and isothermal $\rho_{[100]}(H)$ data reveals that the evolution of electrical resistivity in the neighborhood of the TCP does not correlate with the AHE. A possible explanation may be related to the important role of field-induced spin-flip fluctuations from the AFM state in enhancement of the electrical resistivity at temperatures above $T_{tcp}$.

The first-order metamagnetic transitions are characterised by the simultaneous appearance of pronounced jumps in magnetization, specific heat, electrical resistivity and Hall resistivity. The latter 3 phenomena are common characteristics of phase transitions involving Fermi surface (FS) reconstruction, which are called Lifshitz transitions[56]. Specifically considering uranium intermetallic antiferromagnets, recently much interest has been attracted by possible Lifshitz transitions in UPt$_2$Si$_2$[57,58]. An interesting case is represented by UPd$_2$Al$_3$ in which a cascade of Lifshitz transitions is indicated by anomalies in the Seebeck coefficient in the AFM state in fields lower than $H_c$ which is followed by a Lifshitz MT at $H_c$[59].

The Seebeck coefficient:

$$S = -\frac{\pi^2 k_B^2}{3|e|} T \left[\frac{\partial lnN(E)}{\partial E} + \frac{\partial ln\tau(E)}{\partial E}\right]_{E=E_F} \qquad (7),$$

where $N(E)$ is the density of states and $\tau(E)$ is the relaxation time of conduction electrons[60], is closely connected to characteristics of the Fermi surface. An observed sudden change of $S(H)$ provides indication of a possible change of the energy derivative of the density of states at the Fermi level due to a Fermi surface reconstruction connected with the transition.

In Fig. 8, $S(H)$ dependences measured at 15, 20 and 33 K on the UIrSi$_3$ crystal for $\Delta T // c$ are displayed. A clear drop of the value of Seebeck coefficient at $H_c$ is observed when measured at temperatures below $T_{tcp}$, at which the first-order AFM↔PPM metamagnetic transition takes place. This result, in conjunction with the observed simultaneous jumps in $\rho(H)$, $\rho_H(H)$ and $C_p(H)$ dependences (for results measured at 2 K see Fig. 9), suggests that the FOMPT in UIrSi$_3$ is probably a Lifshitz transition, which is characterised by a Fermi surface reconstruction. When considering that the uranium 5f-electron states for UIrSi$_3$ (carrying magnetic moments) are itinerant and can be present at the Fermi surface, some FS reconstruction due to the change of magnetic periodicity by the AFM↔PPM transition can be expected. Measurements of X-ray Magnetic Circular Dichroism (XMCD), de Haas – van Alphen (dHvA) and/or Shubnikov de Haas (SdH) effect directly testing electronic structure in magnetic fields and relevant band-structure calculations are, however, needed to provide decisive arguments in this issue. On the other hand, no drop at $H_c$ but just a narrow valley has been observed in the $S(H)$ dependence when measured at 33 K ($>T_{tcp}$) as a result of the SOMPT.

The temperature dependence of Seebeck coefficient as measured at low temperatures and zero magnetic field is shown in Fig 8 (right panel). It shows a dip at ordering temperature with the presence of the peak centred at 1/2 $T_N$, as expected from the gapping of the Fermi surface below $T_N$ (see e.g. Ref. 61). This leads to a notable difference in the initial slopes for field dependencies at 15 K and at higher temperatures.

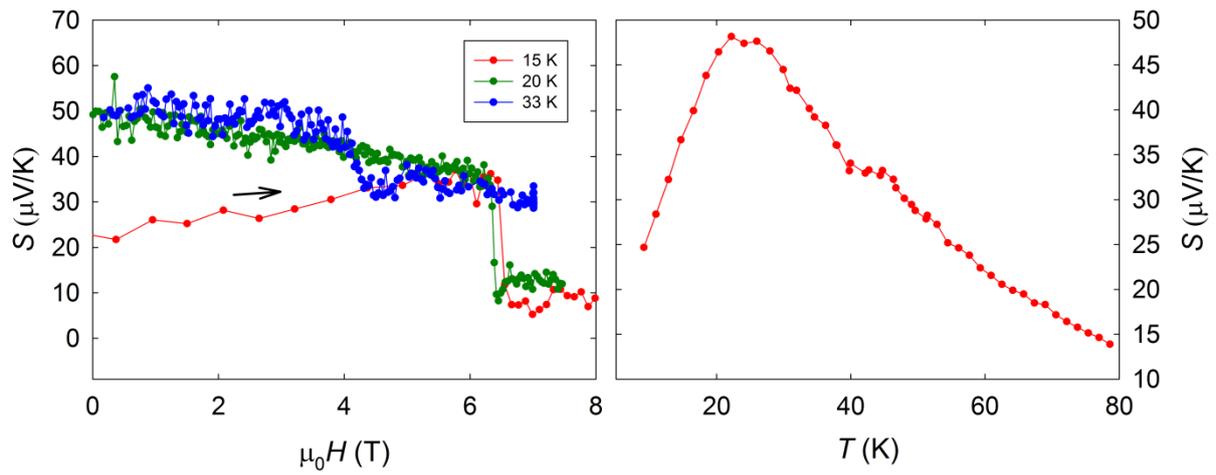

Fig. 8. The Seebeck coefficient of UIrSi$_3$ (left panel) at 15, 20 and 33 K as a function of the magnetic field applied in the [001] direction. The arrow shows the direction of field sweep. Right panel: Temperature dependence in zero magnetic field.

There is one more feature of UIrSi$_3$, which can be seen from the comparison of the $\rho_{[001]}(H)$, $\rho_{[100]}(H)$, $\rho_H(H)$, $M(H)$ and $C_p(H)$ dependences measured at 2 K shown in Fig. 9. One can see that the electrical resistivity indicates an additional field-induced bump between 0 and 4 T, which is, however, not reflected in the field dependences of magnetization, Hall resistance and specific heat, although measured on an identical sample. Assuming a certain analogy with CePtSn[62-64], one can speculate about a transition between two AFM states. Detailed microscopic studies, mainly using neutron scattering and μSR are desired to demonstrate whether the speculation is realistic.

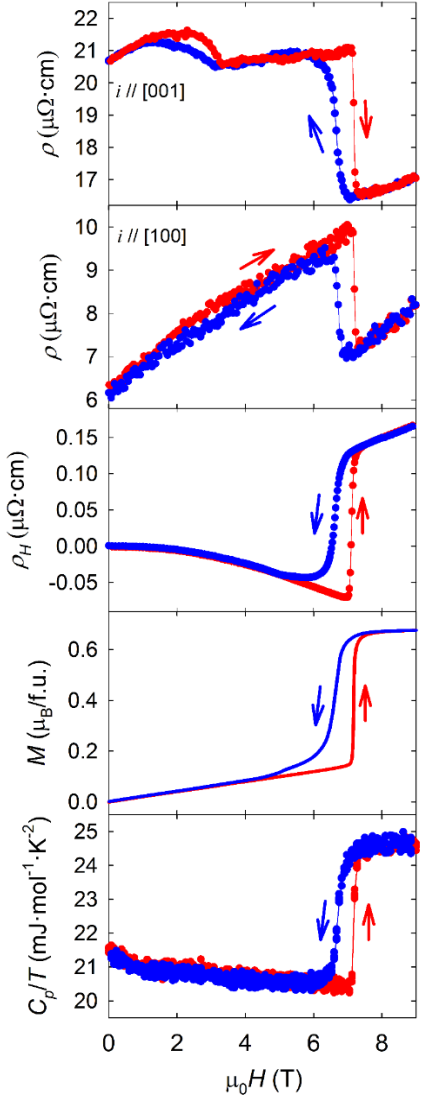

Fig. 9. From top to bottom panel:

- electrical resistivity for $i//[001]$,
- electrical resistivity for $i//[100]$,
- Hall resistivity,
- magnetization,
- specific heat divided by temperature

of UIrSi$_3$ at 2 K as functions of the magnetic field applied in the [001] direction.

In this context it is worth noting that the magnetic contribution to the electrical resistivity carries some information on the magnetic structure as a result of scattering of conduction electrons from magnetic moments in the material. This is to certain extent comparable to the magnetic scattering of neutrons in magnetic materials. The main difference is that the diffraction of neutron flux from an AFM lattice usually provides new magnetic reflections, carrying rich information on magnetic structure, whereas the scattering of conduction electrons provides only a new value of electrical resistivity, providing an indication of a possible change of AFM structure. The sensitivity of the electrical resistance to changes in the magnetic structure of U compounds is enhanced by the strong exchange of conducting electrons with moment-carrying U 5f-electrons, with some states on the Fermi surface.

In Fig. 10, the magnetic phase diagram of UIrSi$_3$ in the magnetic field applied along the $c$-axis is depicted.

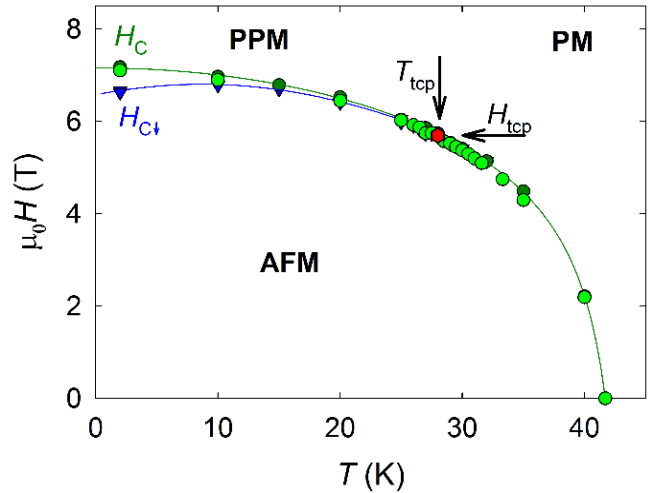

Fig. 10. The magnetic phase diagram of UIrSi$_3$ when the magnetic field is applied along the $c$-axis. PM - normal paramagnet, PPM – polarized paramagnet regime as a result of a FOMPT in fields above $H_c$, AFM – antiferromagnetic phase. $H_c$ and $H_{c\downarrow}$ represented by dark green circles and blue upside-down triangles are defined as inflection points of the field-sweep-up and field-

sweep-down $M(H)$, $\rho(H)$ and $\rho_H(H)$ isotherms, respectively, in the vicinity of the MT. The light green circles represent $T_N$ values determined by anomalies on the isofield $C_p(T)$, $\rho(T)$ and $\rho_H(H)$ curves. The red hexagon represents the tricritical. The lines are guides for eye.

## Conclusions

We have performed a detailed study of the electrical resistance, Hall resistance and thermoelectric power of the Ising non-centrosymmetric antiferromagnet UIrSi$_3$ at various temperatures and magnetic fields with a special emphasis on phenomena associated with magnetic phase transitions between the antiferromagnetic and paramagnetic states. The obtained results demonstrate that the electrical and thermal transport properties can provide valuable information on the character of magnetic phase transformations in antiferromagnets.

We have observed that the unequivocally different character of the FOMPTs and the SOMPTs in UIrSi$_3$ are reflected in the dramatically different transport properties in the neighborhood of the corresponding critical temperatures, $T_N$, and magnetic fields, $H_c$. Considering the magnetic parts of electrical resistivity and Hall resistivity, we have suggested a scenario which may successfully explain the observed change of polarity of the $\Delta\rho_H(T)$ and $\Delta\rho_H(H)$ steps at the TCP which separates the FOMPT and SOMPT segments in the magnetic phase diagram of UIrSi$_3$. Analogous detailed experiments on some other representative Ising-like antiferromagnets are desired to test the universality of the scenario. Neutron-scattering studies of single crystals in magnetic fields are strongly needed in order to confirm the microscopic character of the magnetic regimes assumed in the scenario. Magneto-optic Kerr-effect measurements at various temperatures and magnetic fields would be useful for deeper understanding the underlying mechanism behind the evolution of the AHE in UIrSi$_3$.

The observed simultaneous appearance of pronounced jumps in the field dependence of specific heat, electrical resistivity, Hall resistivity and Seebeck coefficient, respectively, at the FOMPT provide strong indications of a Fermi surface reconstruction, which is characteristic of a magnetic-field induced Lifshitz transition. XMCD, dHvA and/or SdH experiments in cooperation with relevant band-structure calculations are envisaged in order to get more information on the band structure in magnetic fields and test the idea of a Lifshitz transition in UIrSi$_3$.

**Acknowledgments**

This research was supported by the Czech Science Foundation, grant No. 16-06422S and the Japan Society for the Promotion of Science (JSPS) KAKENHI with the grant nos. 15K05156 and 15KK0149. Experiments were performed in the Materials Growth and Measurement Laboratory MGML (http://mgml.eu), which is supported by the Ministry of Education, Youth and Sports within the program of Large Research Infrastructures (grant no. LM2018096). The authors are indebted to Dr. Ross Colman for critical reading and correcting the manuscript.

**Supplementary information**

*Field-sweeping rates*

The field dependences of electrical resistivity $\rho(H)$, Hall resistivity $\rho_H(H)$, thermoelectric power $S(H)$ and magnetization $M(H)$ were measured in fields between 4 and 8 T at a sweep rate of 1 mT/s, 2.5 mT/s, 2.5 mT/s and 2 mT/s, respectively. The system has been found only slightly relaxing. A typical time dependence of electrical resistance at a most "sensitive" point of the hysteresis loop (see Fig. S1) is seen in Fig. S2. This demonstrates that the hysteresis of the FOMPT observed in $\rho(H)$, $\rho_H(H)$, $S(H)$ and $M(H)$ is intrinsic, not an artefact of fast sweeping the applied magnetic field.

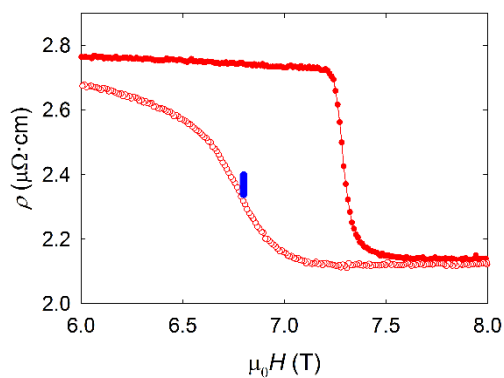

Fig. S1: A part of the field dependence of electrical resistivity of UIrSi$_3$ at 2 K for current parallel to the magnetic field applied along the c-axis (red points). The blue points correspond to the time relaxation dependence shown in Fig. S2.

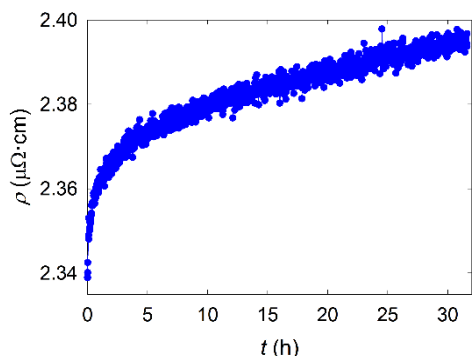

Fig. S2: The time dependence of the electrical resistivity corresponding to blue points in Fig. S1.

*Metamagnetism, PPM regime, PPM ↔ normal PM crossover*

We would like to emphasize that the field-induced (metamagnetic) state of an Ising antiferromagnet in fields above $H_c$ at low temperature is not ferromagnetic (being frequently inadequately quoted in papers) but paramagnetic. The low-temperature PPM regime of the paramagnetic state, though resembling a collinear ferromagnet and having similar impact on transport properties, has a fundamentally different underlying mechanism. A simple collinear ferromagnet is characterized by a spontaneous ordering of magnetic moments coupled by a uniform ferromagnetic interaction. The spontaneous magnetization, which is equal to the saturation magnetization at the low-temperature limit, is the order parameter of a ferromagnet. The polarization of magnetic moments in the PPM regime is reflecting the FM exchange (intra-sublattice) interaction within sublattices of an antiferromagnet (for simplicity a simple two-sublattice AFM is considered), which are coupled antiparallel by the AFM exchange inter-sublattice interaction causing the AFM ground state. When the applied magnetic field reaches $H_c$ the inter-sublattice AFM coupling is overcome by the applied field and the FM-coupled sublattices align parallel. The polarization of moments is assisted by the strong uniaxial magnetocrystalline anisotropy of UIrSi$_3$. The PPM regime of the paramagnetic state is metastable and transforms back to the AFM state as soon as the applied field is reduced below $H_{c\downarrow}$. If the low-temperature field-induced state in our case was ferromagnetic then UIrSi$_3$ should undergo a FM to PM magnetic phase transition when we increase temperature while maintaining the applied magnetic field (e.g. 8 T in our case). Such a transition should be seen on the temperature dependence of specific heat measured in 8 T. The specific-heat data taken in 8 T from 2 to 50 K, however, shows no sign of any anomaly, which would indicate a magnetic phase-transition.

The hierarchy of AFM and FM exchange interactions is almost temperature independent. If the antiferromagnet is cooled from high temperatures in a high magnetic field larger than $H_c$ (say 8 T in the UIrSi$_3$ case) the magnetic moments within sublattices couple ferromagnetically but the antiparallel coupling between the sublattices is prevented by the sufficiently large applied magnetic field. At the same time the FM exchange intra-sublattice interaction polarizes the moments, however, no magnetic phase transition, happens within cooling from high down to the low-temperature limit; the system remains in the paramagnetic state.

A PPM and the normal PM state are not two different magnetic phases but these are just two different regimes of the same paramagnetic phase. These two regimes are therefore not separated by any magnetic phase transition but by a crossover region spreading in the paramagnetic phase space above TCP. This crossover region (see Fig. S3) extends from the low-temperature border $T_{LT}$ (for $T < T_{LT}$ the nearest isofield $M(T)$ curves are parallel indicating the fully polarized regime), and the high-temperature border $T_{HT}$ (for $T > T_{HT}$ the nearest $M/H(T)$ curves coincide within experimental error). Closer inspection of Fig. S3 reveals that the crossover is reflected in the temperature dependence of Hall resistivity mainly as a very broad bump on the negative slope followed by a valley (see Figs. 5). The normal PM regime (PPM regime) seems to be at temperatures above (below) the temperature of the inflection point on the positive (negative) slope of the $\rho_H(T)$ curve above 50 K (below 20 K).

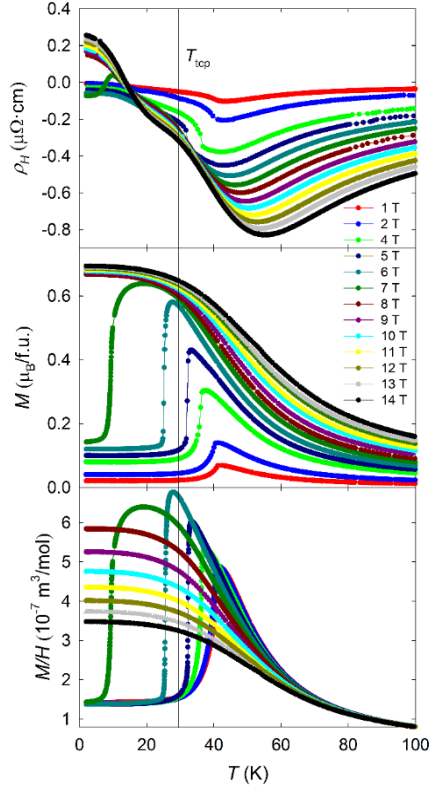

Fig. S3: Temperature dependence of Hall resistivity $\rho_H$ vs. $T$ (top panel), magnetization $M$ vs. $T$ (central panel) and susceptibility magnetization $M/H$ vs. $T$ (bottom panel) of UIrSi$_3$ in magnetic fields in the interval 8 to 14 T applied in the [001] direction. The $\rho_H$ vs. $T$ and $M$ vs. $T$ plots in corresponding magnetic fields are in the same colors. The plots in the top panel are vertically shifted by - 0.05 μΩcm.

*Hall-effect analysis*

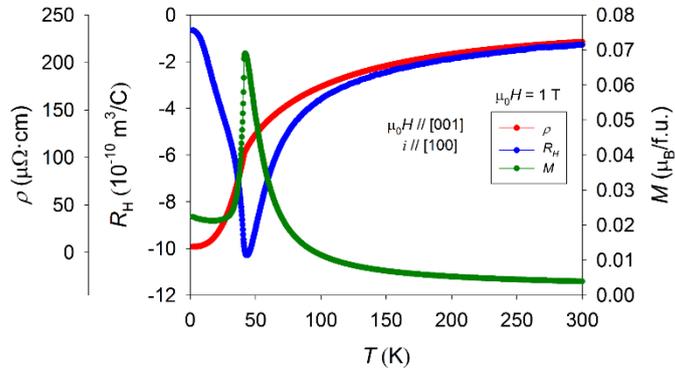

Fig. S4: Temperature dependence of magnetization $M$, Hall coefficient $R_H$, and electrical resistivity $\rho_{[100]}$ of UIrSi$_3$ measured in a field of 1 T applied in the [001] direction.

To investigate the characteristics of magnetic scattering, we have analyzed the AHE. Fig. S4 shows the temperature dependences of the magnetization $M$, Hall coefficient $R_H$ and electrical resistivity for $i$ // [100] measured on UIrSi$_3$ in the 1-T magnetic field // [001] which exhibit clear anomalies at $T_N$. We can see a usual maximum on the $M(T)$ curve and change of slope of $\rho_{[100]}$. $R_H$ is negative within the entire temperature interval 2 -300 K and show a clear minimum in vicinity of the AFM ↔ PM transition.

Usually, only the skew scattering contributes to the AHE in a paramagnetic state, while both the skew scattering and side-jump scattering contribute in the ferromagnetic state[22]. If only the skew scattering contributes to the AHE in UIrSi$_3$ in the paramagnetic state, $R_H$ is expected to be a straight line as a function of $\rho \cdot M$ according formula (4). $R_H$ plotted against $\rho \cdot M$ is shown in Fig. S5, using the data from Fig. S4. We can see that in our case the $R_H(\rho \cdot M)$ plot is not linear at all. Therefore, most probably both, the skew scattering and side-jump scattering contribute to the AHE in UIrSi$_3$.

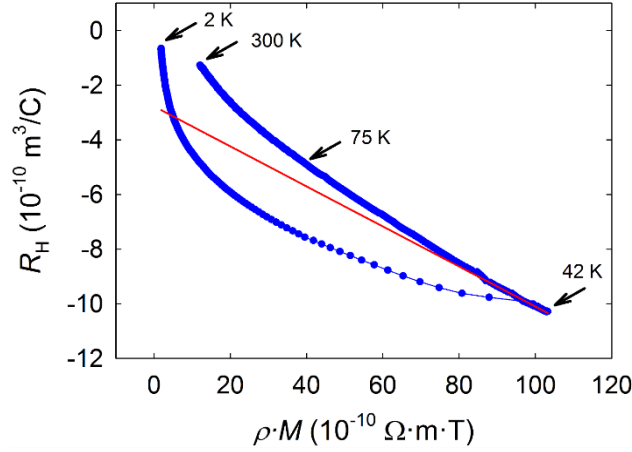

Fig. S5: $R_H$ vs. $\rho_{[100]} \cdot M$ plot for UIrSi$_3$ in the field of 1 T applied in the [001] direction. For details see the text.

As the next step we attempted fitting $R_H$ data obtained in 1 T to the formula:

$$R_H = R_0 + (a \cdot \rho + b \cdot \rho^2) \cdot M/\mu_0 \cdot H \quad (S1),$$

obtained by substituting (4) for $R_s$ in (3). The $R_H$ data cannot be acceptably fitted to (S1) within the entire AFM region 2 – 40 K (PM range 50 – 300 K). Reasonable fits can be obtained within partial AFM segments 2 -17, 17 - 40 K, respectively, and PM segments 50 - 100 K and 100 - 300 K, respectively. The fits are graphically represented in Figs. S6 – S10 and the fitting parameters listed in tables attached to figures and summarized in Tables S7 and S8.

|   | Value | SE | Unit |
|---|---|---|---|
| $R_0$ | $2.91 \cdot 10^{-10}$ | $0.08 \cdot 10^{-10}$ | m$^3$C$^{-1}$ |
| $a$ | -2.41 | 0.06 | T$^{-1}$ |
| $b$ | $2.97 \cdot 10^6$ | $0.3 \cdot 10^6$ | Ωm$^{-1}$T$^{-1}$ |

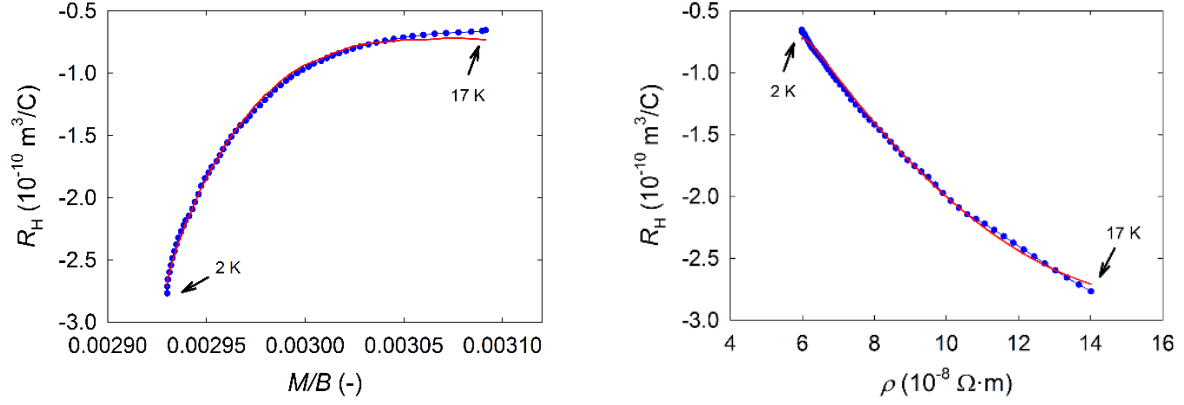

Fig. S6: Results of fitting the Hall coefficient $R_H$ data measured in the field of 1 T applied in the [001] direction to formula (S1) in the temperature interval 2 - 17 K shown in the representation $R_H$ vs. $M/(\mu_0 \cdot H)$ (left panel) and in the representation $R_H$ vs. $\rho$ (right panel).

|   | Value | SE | Unit |
|---|---|---|---|
| $R_0$ | $-1.45 \cdot 10^{-10}$ | $0.02 \cdot 10^{-10}$ | $m^3 C^{-1}$ |
| $a$ | $-3.98 \cdot 10^{-1}$ | $0.03 \cdot 10^{-1}$ | $T^{-1}$ |
| $b$ | $2.97 \cdot 10^{5}$ | $0.03 \cdot 10^{5}$ | $\Omega m^{-1} T^{-1}$ |

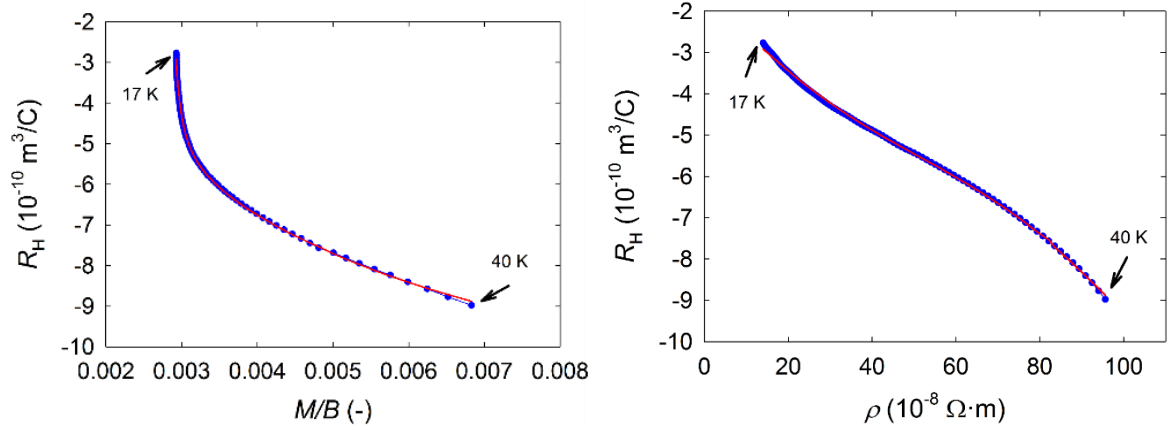

Fig. S7: Results of fitting the Hall coefficient $R_H$ data measured in the field of 1 T applied in the [001] direction to formula (S1) in the temperature interval 17 - 40 K shown in the representation $R_H$ vs. $M/(\mu_0 \cdot H)$ (left panel) and in the representation $R_H$ vs. $\rho$ (right panel).

|   | Value | SE | Unit |
|---|---|---|---|
| $R_0$ | $1.41 \cdot 10^{-11}$ | $0.35 \cdot 10^{-11}$ | $m^3 C^{-1}$ |
| $a$ | $2.82 \cdot 10^{-2}$ | $0.18 \cdot 10^{-2}$ | $T^{-1}$ |
| $b$ | $-6.21 \cdot 10^{4}$ | $0.17 \cdot 10^{5}$ | $\Omega m^{-1} T^{-1}$ |

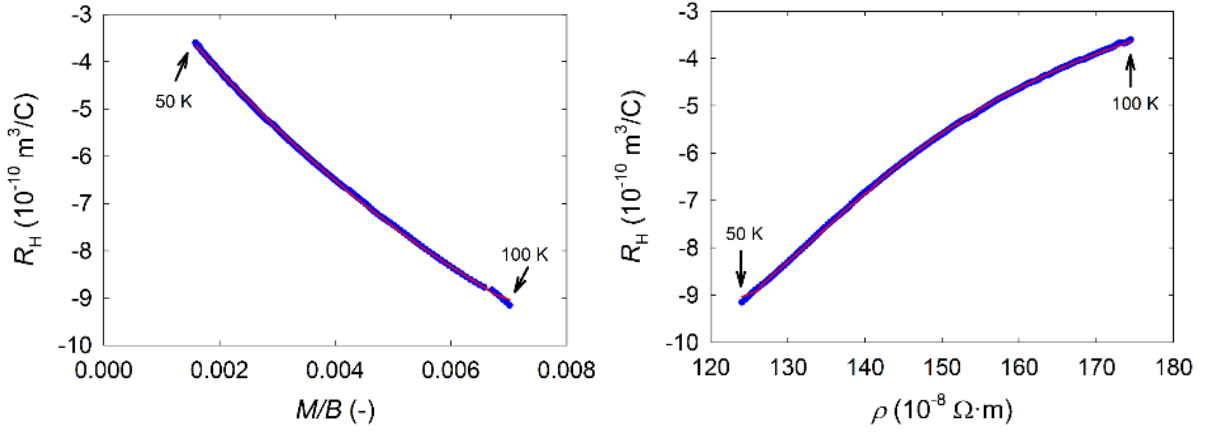

Fig. S8: Results of fitting the Hall coefficient $R_H$ data measured in the field of 1 T applied in the [001] direction to formula (S1) in the temperature interval 50 - 100 K shown in the representation $R_H$ vs. $M/(\mu_0 \cdot H)$ (left panel) and in the representation $R_H$ vs. $\rho$ (right panel).

|   | Value | SE | Unit |
|---|---|---|---|
| $R_0$ | $2.36 \cdot 10^{-10}$ | $0.02 \cdot 10^{-10}$ | $m^3 C^{-1}$ |
| $a$ | $8.58 \cdot 10^{-2}$ | $0.30 \cdot 10^{-2}$ | $T^{-1}$ |
| $b$ | $-1.73 \cdot 10^{5}$ | $0.02 \cdot 10^{5}$ | $\Omega m^{-1} T^{-1}$ |

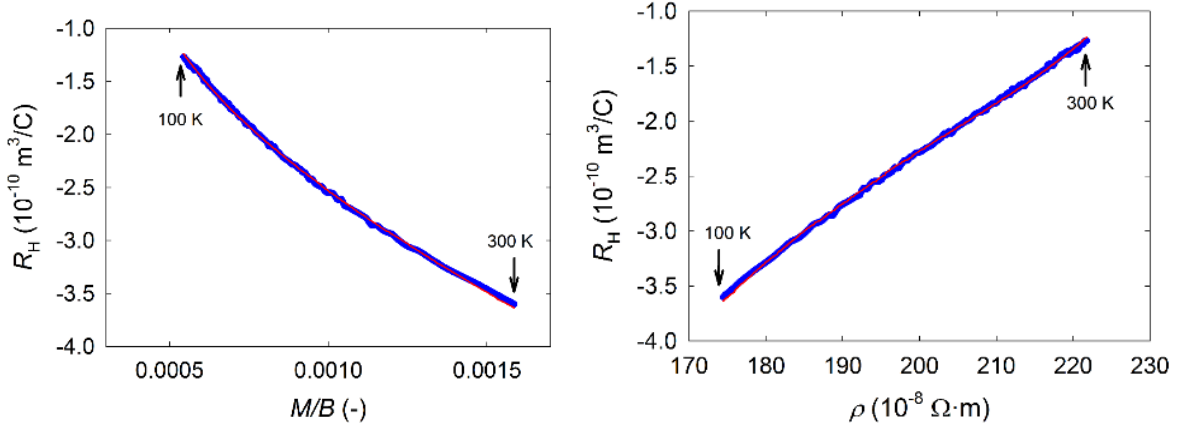

Fig. S9: Results of fitting the Hall coefficient $R_H$ data measured in the field of 1 T applied in the [001] direction to formula (S1) the temperature interval 100 - 300 K shown in the representation $R_H$ vs. $M/(\mu_0 \cdot H)$ (left panel) and in the representation $R_H$ vs. $\rho$ (right panel).

Closer inspection of the fitting parameters indicates that the scenario for the AHE in UIrSi$_3$ is most probably more complex than the empirical approach usually applied to the AHE in ferromagnets. The key evidence seems to be the fact that the fitted values of the ordinary HE coefficient $R_0$ in the temperature intervals 2 -17, 17 - 40 K in AFM state (50 - 100 K and 100 - 300 K in PM state) differ dramatically. Between the 2 -17 K and 17 - 40 K interval we see even a change of sign. Such results do not seem to be physical because one would not expect dramatic changes of electronic structure within the entire AFM (PM) range in low magnetic fields leading to only subtle changes of $R_0$. Also the evolution of the skew- and side-jump-scattering contributions $a\rho$ and $b\rho^2$ obtained from fits is arbitrary within the same phase magnetic (AFM, PM). The side-jump scattering obtained from fitting seems to dominate in the PM state contrary to general observations quoted in literature.

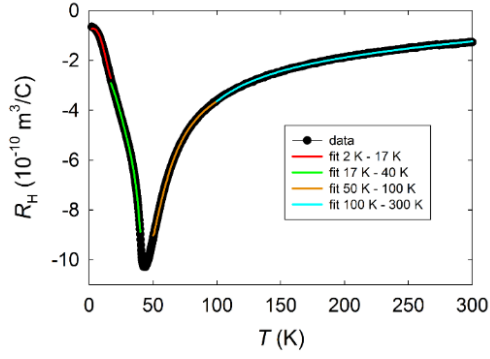

Fig. S10: Results of fitting of Hall coefficient $R_H$ data measured in the field of 1 T applied in the [001] direction to formula (S1) in the temperature intervals 2 -17, 17 - 40 K, 50 - 100 K and 100 - 300 K, respectively, shown in the representation $R_H$ vs. $T$.

TABLE S1. Results of fitting the Hall coefficient $R_H$ measured at 1 T to the formula (S1) in the temperature intervals 2 -17, 17 - 40 K, 50 - 100 K and 100 - 300 K, respectively. The fitting parameters (values in italics) are identical with values given in partial table in Figs. 6-9.

| $T$ | 2-17 | 17-40 | 50 - 100 | 100 - 300 | K |
|---|---|---|---|---|---|
| $R_0$ | $2.91 \cdot 10^{-10}$ | $-1.45 \cdot 10^{-10}$ | $1.41 \cdot 10^{-11}$ | $2.36 \cdot 10^{-10}$ | $m^3 \cdot C^{-1}$ |
| $a$ | $-2.41$ | $-3.98 \cdot 10^{-1}$ | $-2.83 \cdot 10^{-2}$ | $8.58 \cdot 10^{-2}$ | $T^{-1}$ |
| $b$ | $7.46 \cdot 10^6$ | $2.97 \cdot 10^5$ | $-6.21 \cdot 10^4$ | $-1.73 \cdot 10^5$ | $\Omega \cdot m^{-1} \cdot T^{-1}$ |
| $T$ | 2 | 17 | 50 | 100 | K |
| $\rho$ | $5.99 \cdot 10^{-8}$ | $1.40 \cdot 10^{-7}$ | $1.24 \cdot 10^{-6}$ | $1.74 \cdot 10^{-6}$ | $\Omega \cdot m$ |
| $a\rho$ | $-1.44 \cdot 10^{-7}$ | $-5.58 \cdot 10^{-8}$ | $-3.51 \cdot 10^{-8}$ | $1.50 \cdot 10^{-7}$ | $\Omega \cdot m \cdot T^{-1}$ |
| $b\rho^2$ | $2.67 \cdot 10^{-8}$ | $5.84 \cdot 10^{-9}$ | $-9.55 \cdot 10^{-8}$ | $-5.28 \cdot 10^{-7}$ | $\Omega \cdot m \cdot T^{-1}$ |
| $M$ | $3.09 \cdot 10^{-3}$ | $2.93 \cdot 10^{-3}$ | $7.02 \cdot 10^{-3}$ | $1.59 \cdot 10^{-3}$ | T |
| $R_s$ | $-1.18 \cdot 10^{-7}$ | $-4.99 \cdot 10^{-8}$ | $-1.31 \cdot 10^{-7}$ | $-3.78 \cdot 10^{-7}$ | $\Omega \cdot m \cdot T^{-1}$ |
| $R_H$ | $-7.33 \cdot 10^{-11}$ | $-2.92 \cdot 10^{-10}$ | $-9.02 \cdot 10^{-10}$ | $-3.64 \cdot 10^{-10}$ | $m^3 \cdot C^{-1}$ |
| $|R_0/R_H|$ | 3.97 | 0.50 | 0.02 | 0.65 | - |
| $R_0/(R_H-R_0)$ | $-0.80$ | 0.99 | $-0.02$ | $-0.39$ | - |
| $T$ | 17 | 40 | 100 | 300 | K |
| $\rho$ | $1.40 \cdot 10^{-7}$ | $9.56 \cdot 10^{-7}$ | $1.74 \cdot 10^{-6}$ | $2.22 \cdot 10^{-6}$ | $\Omega m$ |
| $a\rho$ | $-3.38 \cdot 10^{-7}$ | $-3.81 \cdot 10^{-7}$ | $-4.93 \cdot 10^{-8}$ | $1.90 \cdot 10^{-7}$ | $\Omega \cdot m \cdot T^{-1}$ |
| $b\rho^2$ | $1.46 \cdot 10^{-7}$ | $2.72 \cdot 10^{-7}$ | $-1.89 \cdot 10^{-7}$ | $-8.53 \cdot 10^{-7}$ | $\Omega \cdot m \cdot T^{-1}$ |
| $M$ | $2.93 \cdot 10^{-3}$ | $6.83 \cdot 10^{-3}$ | $1.59 \cdot 10^{-3}$ | $5.44 \cdot 10^{-4}$ | T |
| $R_s$ | $-1.92 \cdot 10^{-7}$ | $-1.09 \cdot 10^{-7}$ | $-2.38 \cdot 10^{-7}$ | $-6.63 \cdot 10^{-7}$ | $\Omega \cdot m \cdot T^{-1}$ |
| $R_H$ | $-2.71 \cdot 10^{-10}$ | $-8.88 \cdot 10^{-10}$ | $-3.64 \cdot 10^{-10}$ | $-1.24 \cdot 10^{-10}$ | $m^3 \cdot C^{-1}$ |
| $|R_0/R_H|$ | 1.07 | 0.16 | $-0.04$ | 1.91 | - |
| $R_0/(R_H-R_0)$ | $-0.52$ | 0.20 | $-0.04$ | $-0.66$ | - |

The AHE measured in ferromagnets in a zero magnetic field represents the contribution due to the spontaneous magnetization. In low magnetic fields the AHE in a FM state is usually orders of magnitude larger than the ordinary HE contribution. The AHE in collinear antiferromagnets in zero field should be zero because of the absence of bulk spontaneous magnetization. The ordinary HE and AHE in UIrSi$_3$ obtained from fitting in the AFM state in 1 T and 4 T at temperatures below 17 K (see Tables S1 and S2) appear to be of the same order of magnitude at low temperatures. At higher temperatures ($\geq 17$ K) the AHE contribution seems to dominate.

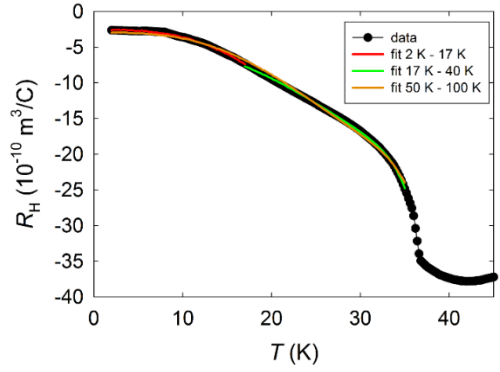

Fig. S11: Results of fitting the Hall coefficient $R_H$ data measured in the field of 4 T applied in the [001] direction to formula (S1) in the temperature intervals 2 -17, 17 - 35 K, 2 - 35 K shown in the representation $R_H$ vs. $T$.

TABLE S2: Results of fitting the Hall coefficient $R_H$ data measured in the field of 4 T to the formula (S1) in the temperature intervals 2 -17, 17 - 35 K, 2 - 35 K, respectively.

| $T$ | 2-17 | | 17-35 | | 2 - 35 | | K |
|---|---|---|---|---|---|---|---|
| | Value | SE | Value | SE | Value | SE | Unit |
| $R_0$ | $3.90 \cdot 10^{-10}$ | $0.27 \cdot 10^{-10}$ | $-3.20 \cdot 10^{-11}$ | $0.84 \cdot 10^{-11}$ | $2.19 \cdot 10^{-10}$ | $0.08 \cdot 10^{-10}$ | m$^3$·C$^{-1}$ |
| $a$ | -3.42 | 0.20 | -2.18 | 0.02 | -2.69 | 0.03 | T$^{-1}$ |
| $b$ | $3.94 \cdot 10^6$ | $0.95 \cdot 10^6$ | $2.16 \cdot 10^6$ | $0.01 \cdot 10^6$ | $2.86 \cdot 10^6$ | $0.04 \cdot 10^6$ | Ω·m$^{-1}$·T$^{-1}$ |
| $T$ | 2 | | 17 | | 2 | | K |
| $\rho$ | $7.69 \cdot 10^{-8}$ | | $1.45 \cdot 10^{-7}$ | | $7.69 \cdot 10^{-8}$ | | Ω·m |
| $a\rho$ | $-2.63 \cdot 10^{-7}$ | | $-3.16 \cdot 10^{-7}$ | | $-2.07 \cdot 10^{-7}$ | | Ω·m·T$^{-1}$ |
| $b\rho^2$ | $2.33 \cdot 10^{-8}$ | | $4.56 \cdot 10^{-8}$ | | $1.69 \cdot 10^{-8}$ | | Ω·m·T$^{-1}$ |
| $M$ | $2.72 \cdot 10^{-3}$ | | $2.73 \cdot 10^{-3}$ | | $2.72 \cdot 10^{-3}$ | | T |
| $R_s$ | $-2.40 \cdot 10^{-7}$ | | $-2.70 \cdot 10^{-7}$ | | $-1.90 \cdot 10^{-7}$ | | Ω·m·T$^{-1}$ |

| | | | | |
|---|---|---|---|---|
| $R_H$ | $-2.62 \cdot 10^{-10}$ | $-7.70 \cdot 10^{-10}$ | $-2.99 \cdot 10^{-10}$ | $m^3 \cdot C^{-1}$ |
| $|R_0/R_H|$ | $-1.49$ | $0.04$ | $-0.73$ | - |
| $R_0/(R_H-R_0)$ | $-0.60$ | $0.04$ | $-0.42$ | - |
| $T$ | $17$ | $35$ | $35$ | K |
| $\rho$ | $1.45 \cdot 10^{-7}$ | $6.65 \cdot 10^{-7}$ | $6.65 \cdot 10^{-7}$ | $\Omega m$ |
| $a\rho$ | $-4.97 \cdot 10^{-7}$ | $-1.45 \cdot 10^{-6}$ | $-1.79 \cdot 10^{-6}$ | $\Omega \cdot m \cdot T^{-1}$ |
| $b\rho^2$ | $8.31 \cdot 10^{-8}$ | $9.57 \cdot 10^{-7}$ | $1.27 \cdot 10^{-6}$ | $\Omega \cdot m \cdot T^{-1}$ |
| $M$ | $2.73 \cdot 10^{-3}$ | $4.98 \cdot 10^{-3}$ | $4.98 \cdot 10^{-3}$ | T |
| $R_s$ | $-4.14 \cdot 10^{-7}$ | $-4.91 \cdot 10^{-7}$ | $-5.26 \cdot 10^{-7}$ | $\Omega \cdot m \cdot T^{-1}$ |
| $R_H$ | $-7.39 \cdot 10^{-10}$ | $-2.48 \cdot 10^{-9}$ | $-2.40 \cdot 10^{-9}$ | $m^3 \cdot C^{-1}$ |
| $|R_0/R_H|$ | $-0.53$ | $0.01$ | $-0.09$ | - |
| $R_0/(R_H-R_0)$ | $-0.35$ | $0.01$ | $-0.08$ | - |

We also fitted the field dependences of Hall resistivity (see Fig. S12, S13) to formula:

$$\rho_H = R_0 \cdot \mu_0 \cdot H + (a \cdot \rho + b \cdot \rho^2) \cdot M \quad (S2),$$

For 2-K data in fields 0 – 4 T we have received a good fit with $R_0 < 0$, $a > 0$ and $b < 0$ (see Table S3). Calculating $\rho_H(H)$ at 2 K beyond 4 T using the above mentioned fitted parameters we have received reasonable agreement of calculated $\rho_H(H)$ values with experimental data in field almost up to $H_c$ and a sharp positive jump at $H_c$, which was, however, much larger than measured. Supposing that the normal Hall coefficient is to be invariable in the AFM state, we have fixed the value $R_0 = -1.79 \cdot 10^{-10}$ m$^3$/C obtained by fitting 2-K data for fitting the $\rho_H(H)$ at other temperatures. The fit of 10-K data was still reasonable including the positive $\Delta \rho_H$ jump at $H_c$.

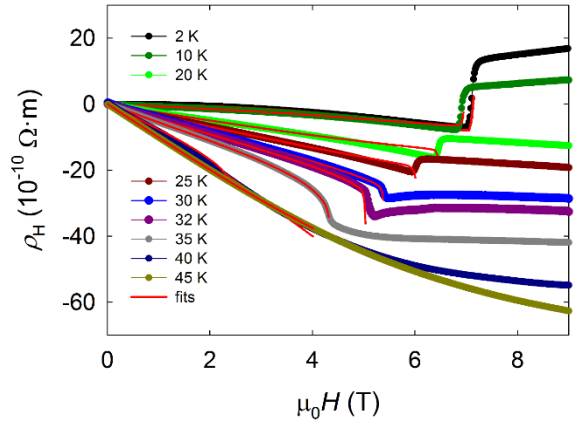

Fig. S12: Results of fitting the Hall resistivity $\rho_H$ data measured at various temperatures as functions of magnetic field applied in the [001] direction to formula (S2) with a fixed value of $R_0 = -1.79 \cdot 10^{-10}$ m$^3$/C in the field interval 0 – 4 T shown in the representation $\rho_H$ vs. $\mu_0 \cdot H$.

Table S3: Results of fitting the Hall resistivity $\rho_H$ data measured at various temperatures as functions of magnetic field applied in the [001] direction to formula (S2 with a fixed value of $R_0 = -1.79 \cdot 10^{-10}$ m$^3$/C) in the field interval 0 – 4 T.

| $T$ (K) | $a$ (T$^{-1}$) | | $b$ ($\Omega\cdot$m$^{-1}\cdot$T$^{-1}$) | |
|---|---|---|---|---|
| | Value | SE | Value | SE |
| 2 | 2.43 | 0.12 | $-2.31\cdot 10^7$ | $0.08\cdot 10^7$ |
| 10 | 1.20 | 0.02 | $-9.04\cdot 10^6$ | $0.26\cdot 10^6$ |
| 20 | $-6.74\cdot 10^{-1}$ | $0.19\cdot 10^{-1}$ | $3.03\cdot 10^6$ | $0.11\cdot 10^6$ |
| 25 | $-8.06\cdot 10^{-1}$ | $0.29\cdot 10^{-1}$ | $2.16\cdot 10^6$ | $0.10\cdot 10^6$ |
| 30 | $-2.93\cdot 10^{-1}$ | $0.16\cdot 10^{-1}$ | $2.42\cdot 10^5$ | $0.38\cdot 10^5$ |
| 32 | $-3.67\cdot 10^{-1}$ | $0.02\cdot 10^{-1}$ | $3.85\cdot 10^5$ | $0.05\cdot 10^5$ |
| 35 | $-4.21\cdot 10^{-1}$ | $0.02\cdot 10^{-1}$ | $4.27\cdot 10^5$ | $0.03\cdot 10^5$ |
| 40 | $-3.36\cdot 10^{-1}$ | $0.34\cdot 10^{-1}$ | $2.79\cdot 10^5$ | $0.39\cdot 10^5$ |
| 45 | $-3.34\cdot 10^{-2}$ | $0.08\cdot 10^{-2}$ | $-6.57\cdot 10^4$ | $0.08\cdot 10^4$ |

The fitted parameters $a$ and $b$ changed values, however, kept their signs ($a > 0$ and $b < 0$). For higher temperatures $a$ and $b$ swapped sign, the calculated $\Delta\rho_H$ at $H_c$ became negative which is opposite to the measured positive jump at $T < T_{tcp}$. At $T > T_{tcp}$ the calculated and measured $\Delta\rho_H$ have the same sign but the huge quantitative disagreement indicated that the fit is probably far from reality.

As the next step we fitted the $\rho_H(H)$ dependences to formula (S2 with all three parameters $R_0$, $a$ and $b$ free) in the field interval 0 T – $H_c$. The fits were reasonable almost up to $H_c$ and also the $\Delta\rho_H$ jumps at $H_c$ were qualitatively (including polarity) reproduced (see Fig. S13). The values fitting parameters were considerably varying with respect to temperature, nevertheless their signs ($R_0 < 0$, $a > 0$ and $b < 0$) were maintained (see Table S4).

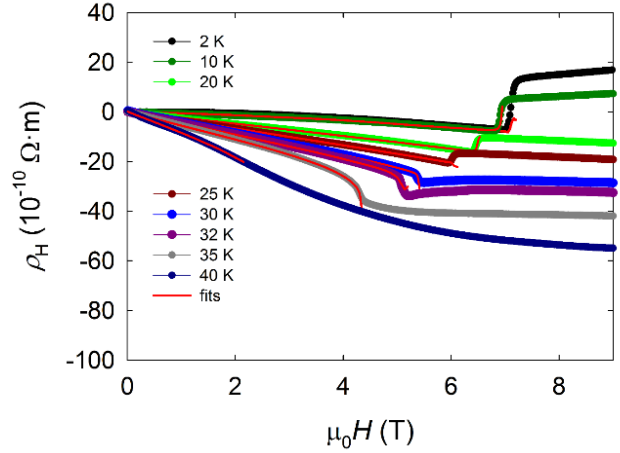

Fig. S13: Results of fitting the Hall resistivity $\rho_H$ data measured at various temperatures as functions of magnetic field applied in the [001] direction to formula (S2 with free $R_0$, $a$, $b$ parameters) in the field interval 0 T – $H_c$ shown in the representation $\rho_H$ vs. $\mu_0\cdot H$.

Table S4: Results of fitting the Hall resistivity $\rho_H$ data measured at various temperatures as functions of magnetic field applied in the [001] direction to formula (S2 with free $R_0$, $a$, $b$ parameters) in the field interval 0 T – $H_c$ shown in the representation $\rho_H$ vs. $\mu_0\cdot H$.

| $T$ (K) | $R_0$ (m³·C⁻¹) | | $a$ (T⁻¹) | | $b$ (Ω·m⁻¹·T⁻¹) | |
| --- | --- | --- | --- | --- | --- | --- |
| | Value | SE | Value | SE | Value | SE |
| 2  | -6.13·10⁻¹⁰ | 0.07·10⁻¹⁰ | 5.94  | 0.06 | -4.24·10⁷ | 0.03·10⁷ |
| 10 | -3.25·10⁻¹⁰ | 0.19·10⁻¹⁰ | 2.77  | 0.15 | -2.00·10⁷ | 0.09·10⁷ |
| 20 | -2.59·10⁻¹⁰ | 0.02·10⁻¹⁰ | 0.19  | 0.01 | -8.66·10⁵ | 0.60·10⁵ |
| 25 | -3.21·10⁻¹⁰ | 0.09·10⁻¹⁰ | 0.04  | 0.03 | -2.05·10⁵ | 0.78·10⁵ |
| 30 | -3.89·10⁻¹⁰ | 0.03·10⁻¹⁰ | 0.19  | 0.01 | -5.26·10⁵ | 0.22·10⁵ |
| 32 | -5.28·10⁻¹⁰ | 0.03·10⁻¹⁰ | 0.36  | 0.01 | -6.40·10⁵ | 0.17·10⁵ |
| 35 | -2.14·10⁻¹⁰ | 0.14·10⁻¹⁰ | -0.36 | 0.02 |  3.55·10⁵ | 0.30·10⁵ |
| 40 |  2.72·10⁻¹⁰ | 1.60·10⁻¹⁰ | -1.00 | 0.19 |  9.48·10⁵ | 1.89·10⁵ |

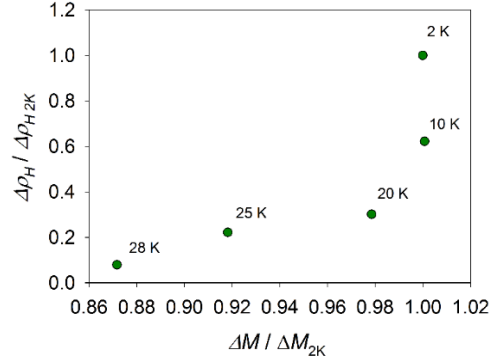

Fig. S14: Relation of jumps of $\Delta\rho_H$ vs. $\Delta M$ at the FOMPT normalized to the corresponding 2-K values.

Fig. S14 demonstrates that the jump of the Hall resistivity $\Delta\rho_H$ at FOMPT is far from scaling with the corresponding magnetization jump $\Delta M$. When increasing temperature from 2 to 20 K $\Delta\rho_H$ decreases by 70 % while $\Delta M$ is reduced by only 2 %. Whereas $\Delta M$ still maintains 87% of its low temperature value, in close vicinity of the TCP $\Delta\rho_H$ vanishes.

*Change of polarity of $\Delta\rho_H(T)$ and $\Delta\rho_H(H)$ at TCP*

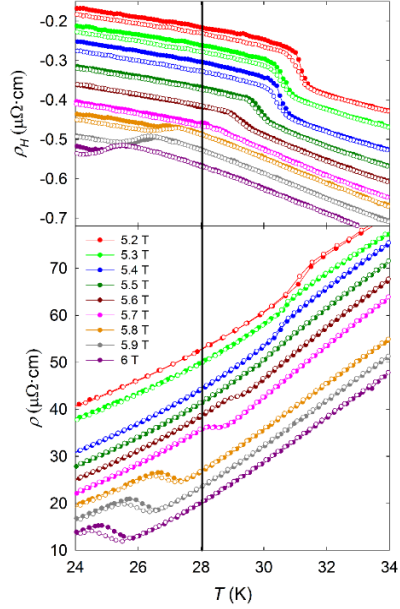

Fig. S15: Temperature dependences of Hall resistivity (upper panel) and electrical resistivity (lower panel) of UIrSi$_3$ measured in selected magnetic fields (applied in the [001] direction) in the vicinity of the TCP. Only data collected with increasing temperature are displayed. The vertical line represents an estimation of the point ($T \sim 28$ K for $\mu_0H \sim 5.7$ T) at which $\Delta\rho_H(T)$ values change polarity.

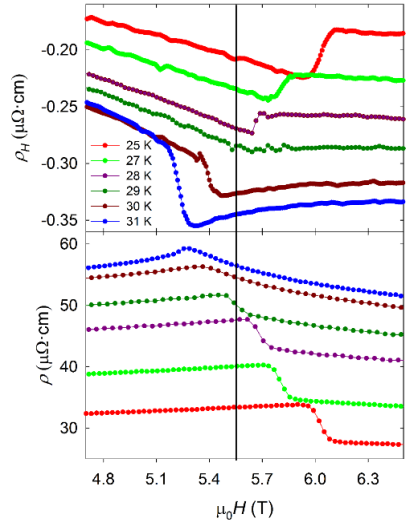

Fig. S16: The Hall resistivity (upper panel) and electrical resistivity (lower panel) of UIrSi$_3$ at selected temperatures in the vicinity of the TCP as functions of the magnetic field applied in the [001] direction. Only data collected with increasing temperature are displayed. The vertical line represents estimation of the point at which $\Delta\rho_H(H)$ values change polarity ($\mu_0H \sim 5.6$ T at $T \sim 29$ K).

In order to explore the details of the evolution of the $\rho_H$ and $\rho_{[100]}$ anomalies in the neighborhood of the TCP in the *T-H* phase space we performed thorough measurements of $\rho_H(T)$ and $\rho_{[100]}(T)$ isofield curves for fields 5.2, 5.3, …5.9, 6.0 T and ($\rho_H(H)$ and $\rho_{[100]}(H)$ isotherms at temperatures 25, 26, …, 30, 31 K). The results of these measurements are displayed in Figs. S15 and S16. It is evident that $\Delta\rho_H(T)$ and $\Delta\rho_H(H)$ continually develop from positive to negative values with decreasing magnetic field and increasing temperature, respectively. Considering the estimated values of temperatures and fields for which $\Delta\rho_H(T)$ and $\Delta\rho_H(H)$ values pass through zero we conclude that the change of polarity of jumps of the AHE as functions of temperature and magnetic field takes place at the TCP.

Closer inspection of isofield $\rho_{[100]}(T)$ and isothermal $\rho_{[100]}(H)$ data reveals that the evolution of electrical resistivity in the neighborhood of the TCP does not correlate with the AHE. A possible explanation may be related to the important role of field-induced spin-flip fluctuations from the AFM state in enhancement of electrical resistivity at temperatures above $T_{tcp}$.